# A discussion on "generality of shear thickening in dense suspensions"

Yu Tian*, Yonggang Meng

State Key Laboratory of Tribology, Tsinghua University, Beijing 100084, P. R. China

## Abstract

In a recent Nature Materials article, Brown et al. reported a generality of shear thickening in dense suspensions and demonstrated that shear thickening can be masked by a yield stress and can be recovered when the yield stress is decreased to below a threshold. However, the generality of the shear thickening reported in the article may not be necessary true when a high electric/magnetic field is applied on an ER/MR fluid. Shear thickening of ER fluid and MR fluid under high electric/magnetic fields at low shear rates, indicating an obvious phase change inside the dense suspensions has been observed.

Correspondence E-mail: tianyu@tsinghua.edu.cn

In a recent Nature Materials article, Brown et al. reported a generality of shear thickening in dense suspensions and demonstrated that shear thickening can be masked by a yield stress and can be recovered when the yield stress is decreased to below a threshold [1]. In our view, the generality of the shear thickening reported in the article may not be always true. It could not incorporate the shear thickening of ER fluid and MR fluid under high electric/magnetic fields, which has been claimed to not happen by Brown et al. Typical shear curves of an ER fluid under various electric fields are shown in Fig. 1. When the applied electric field is below 900 V/mm, only shear thinning is observed, being consistent with the statement of Brown et al. However, when the applied electric field reaches to or exceeds 900 V/mm, a "discontinuous" shear thickening is observed with a sudden shear stress increase at certain critical shear rates. These shear thickenings have been verified by various tests of ER fluids with different shear stress, shear rate, or electric field control modes [2]. The results also indicate that shear thickening plays important roles in ER effect to obtain a high shear yield stress. Similar shear thickening phenomenon has also been found in MR fluids at high shear stress and high magnetic fields [3]. In our view, the rheological property and shear thickening of anisotropic dipolar suspension is greatly different from that of isotropic attractive suspensions. The generality of the above article could not be extended to the area of ER/MR suspensions or the ER/MR effect.

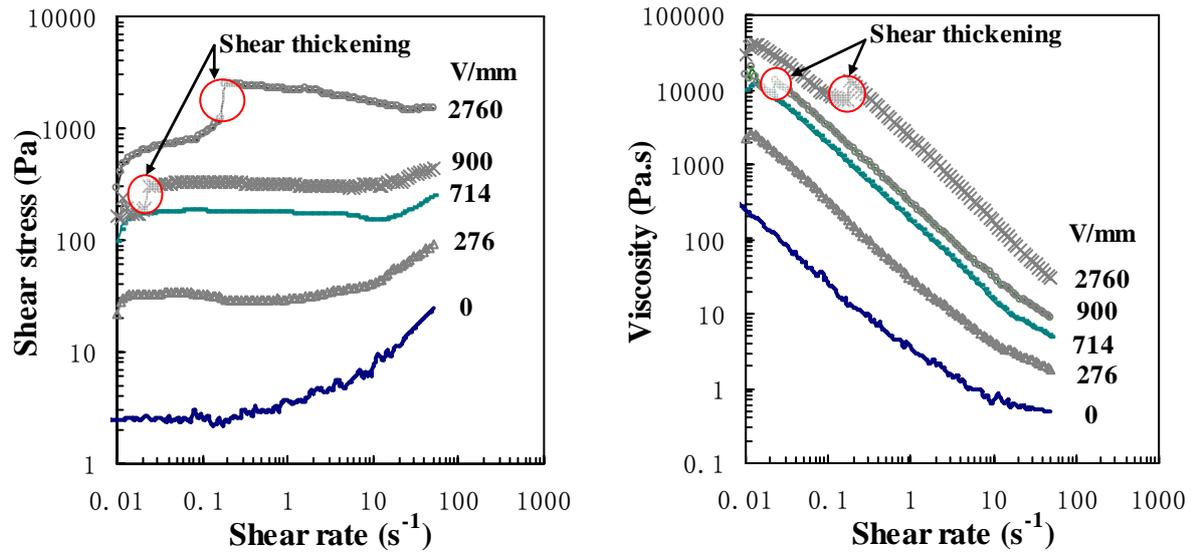

Fig. 1 The rheological property of an ER fluid based on zeolite and silicone oil with a particle volume fraction of 20% and tested on a Anton Par MCR 301 rheometer between two concentric cylinders. (a) Curves of shear stress versus shear rate under no electric field or a certain electric field during the whole shearing process. (b) The viscosity change of the ER fluid sheared under different electric fields.

On the other hand, ER and MR effects are generally characterized by their significant apparent viscosity increase and the forming of dipolar chain structures inside the ER/MR suspensions upon the applying of external electric/magnetic fields [4-6]. The shear yield stresses of ER/MR suspensions under high fields are generally on the order of kPa under several kV/mm or hundreds of mT. Both the shear stress and fields are much higher than those represented in the above article. The traditional polarization model used in the article to calculate the yield stress of ER suspension has been widely recognized to give a much lower estimation than attainable values of real ER fluids [5]. In our view, the slight rheological property change shown in the above article should not be termed as ER/MR effect or ER/MR yield stress.

Former researches have found that the onset of shear thickening is mainly governed by the competitions among viscous force, particle interaction, and the thermal Brownian force [7-8]. Different liquid mediums of a suspension with different viscosities would change the viscous force under the same shear rate. Liquids with different dielectric properties would lead to different particle interactions due to DLVO forces [9]. The term of "particle-liquid surface tension" has been utilized to elucidate the clustering of particles. But this term only refers to the surface tension at the interface of particle-liquid, which does not drive particles to aggregate. Also, the authors state "In the aqueous environment the coating leads to network-like particle clusters (Fig. 1b, top), which minimize exposed surface area and thus potential energy." It is well-known in colloid and surface science, when particle interactions are attractive, particles tend to aggregate or cluster to minimize the free energy of the system [10]. Strictly, even when particles already aggregated, the real surface area of particles exposed to the continuous medium may not obviously change. In a more general case, the continuous medium could be confined in the contact region and acts as a boundary lubrication film or thin lubrication film [11]. Thus, the description of "the minimization of the exposed surface area and thus potential energy" should be more careful.